\documentclass[preprintnumbers, prd, onecolumn, floatfix, superscriptaddress, nofootinbib]{revtex4-1}
\usepackage{graphicx}
\usepackage{epsfig}
\usepackage{bm}
\usepackage{amssymb}
\usepackage{float}
\usepackage{amsmath}
\usepackage{subfigure}
\usepackage{dcolumn}
\usepackage{cancel}
\usepackage[colorlinks]{hyperref}
\usepackage[usenames,dvipsnames]{color}
\usepackage{enumitem}

\hypersetup{
breaklinks=true,
pdfstartview={FitH},  
colorlinks=true,      
linkcolor=blue,       
citecolor=red,    
filecolor=magenta,    
urlcolor=blue,        
anchorcolor=green,   
linktocpage=true
}

\def\doi{http://doi.org}

\begin{document}

\title{ Curvature dominance DE-model in f(R)-gravity}
\author{G. K. Goswami}
\email{gk.goswami9@gmail.com}
\affiliation{Department of Mathematics, Netaji Subhas University of Technology, New Delhi-110 078, India}
\author{Rita Rani}
\email{rita.ma19@nsut.ac.in}
\affiliation{Department of Mathematics, Netaji Subhas University of Technology, New Delhi-110 078, India}
\author{Harshna Balhara}
\email{harshna.ma19@nsut.ac.in}
\affiliation{Department of Mathematics, Netaji Subhas University of Technology, New Delhi-110 078, India}
\author{J. K. Singh}
\email{jksingh@nsut.ac.in}
\affiliation{Department of Mathematics, Netaji Subhas University of Technology, New Delhi-110 078, India}

\begin{abstract}

We have probed a cosmological model in $ f(R) $-gravity, which is a cubic equation in scalar curvature $ R $. The  terms arise due to nonlinear $ f(R) $ function are treated as energy due to curvature inspired geometry. As a result, we find accelerating expansion in the universe, which creates an anti-gravitating negative pressure in it. Some of the  physical parameters are solved using numerical methods. The evolution of the model are examined by the latest observational Hubble data (46-data points) and Pantheon data (the latest compilation of SNIa with 40 binned in the redshift range $ 0.014 \leqslant z \leqslant1.62 $). Some important features of the model have been discussed by analyzing the plots of various dynamical parameters. The plots of deceleration parameter $ q $ and the Hubble parameter $ H $ describe the accelerating expansion in the evolution of the Universe at the present epoch. The transition from deceleration to acceleration for our model is obtained at redshift $ z_{tr} \simeq 0.694069 $, which is in good agreement with $ \Lambda $CDM. We have also carried out state finder analysis for our model. The analysis of specific features of the model confirms that our model is consistent with $ \Lambda $CDM in late times.
\end{abstract}

\maketitle

\section{Introduction}  

\qquad A study of origin, evolution and the large scale structure of the universe is the subject of cosmology. It is a well established fact that our universe is spatially homogeneous and isotropic (SHI) at large scale in the unit of $ Mega parsecs $ and $ Giga \, years $. A veteran Friedman-Lema$ \hat{i} $tre-Robertson-Walker $ (FLRW) $  space-time represents the SHI universe. When Einstein Field Equations are solved for FLRW space-time, they describe an expanding and decelerating universe which means that the various galaxies present in it are moving away from us at a slowly decreasing pace with time. The present day surveys and observation \cite{ref1,ref2,ref3,ref4,ref5} suggest a different type of scenario which tell us that  the various galaxies are moving away from us at a relatively increasing pace with time. The literatures tell us that cold dark matters of gravitating origin and dark energy of repulsive anti gravitating origin are present in abundance in our universe which are responsible for the change in the mode of the universe.\\

The cosmological constant once introduced by Einstein in his field equations to describe a physical static universe was resurrected in order to explain acceleration in it known as $ \Lambda $CDM model, which fits best on observational ground \cite {ref6, ref7, ref8, ref9, ref10, ref11}. Despite, it fails in explaining fine-tuning that vacuum energy is very small compared to typical particle physics scales \cite {ref12}. Cosmological tracking solutions  \cite{ref13, ref14} were proposed to overcome the issues against the $ \Lambda $CDM. By that time, a different way of thinking was developed. It was thought that Dark energy must arise from gravitational origin and some non-linear term of Ricci scalar $ R $ in the form of an arbitrary function $ f(R) $ was introduced in the Einstein Hilbert Action to arrive at an alternative to Einstein theory. The name was given $ f(R) $ gravity. In this theory acceleration is obtained spontaneously from the gravitational sector.\\

In 2003, the $ f(R)= R + \frac{a}{R} + {b}{R^2} $ formalism was first enunciated by Nojiri and Odintsov \cite{ref15} to explain the early inflation and the acceleration in late times. Starobinsky \cite{ref16} used a different form of $ f(R) $ and developed a cosmology without a cosmological constant. Sotiriou and Liberati \cite{ref17, ref18} presented  a Palatini Metric-affine formalism of $ f(R) $ gravity. It is noteworthy to mention the work of Srivastava \cite{ref19} who has developed a model in $ f(R) $ gravity, which represents both early (inflation) and late acceleration in the Universe. Some remarkable works on $ f(R) $-gravity have also been carried out by many authors, which are stated in \cite{ref20, ref21, ref22, ref23, ref24, ref25, ref26}.\\

In this paper, we have developed a model filled with perfect fluid in $ f(R) $ gravity  by taking $ R + a R^2 + {b}{R^3} $ as a particular form of $ f(R) $. In the background FLRW space-time, we have considered two field equations which resemble Einstein field equations (EFE). The additional terms which arise due to non-linear $ f(R) $ function are shifted to the right-hand side of field equations that are treated as energy expressions due to curvature inspired geometry. As a result, they bring acceleration in the universe and put anti-gravitating negative pressure in it. The paper is presented in the following sequences. In section II, we have developed $ f(R) $ field equations for FLRW space-time, then taking a particular case $ f(R)= R + \alpha R^2 +\beta R^3 $, we have found numerical solutions for Hubble, deceleration and jerk parameters. In section III, we have used a data set of $ 46 $ Hubble parameters to estimate present value of matter energy parameter $ (\Omega_m) $, density $ (\rho_m) $ and equation of state (EoS) parameter for curvature $ (\omega_k) $. On the basis of these estimated values, we present Error plots for Hubble parameters and distant modulus. The plots show that our theoretical plots fit well with the observed Hubble and Pantheon data. Moreover, the plots match at par with the $ \Lambda $CDM model. The deceleration parameter $ q $ versus redshift $ z $ plot describes the accelerating universe at the present epoch. The red shift transition $ z_t\simeq 0.694069 $ in our model causes transient acceleration in early and late times, which is in good agreement with $ \Lambda $CDM. In section III, we have also carried out the state finder analysis for our model. The analysis confirms that our model lies near $ \Lambda $CDM, and it fits well on observational ground. In the last section, we interpret our work for our obtained curvature dominated model.\\
    
\section{ Einstein Field Equations in $ f(R) $-formalism} 

\qquad The $ f(R) $-gravity general action with matter Lagrangian $ S_m $ yields \cite{ref15}
\begin{equation} \label{2}
S= \int\Big(\frac{1}{16\pi G} f(R)+S_m\Big)\sqrt{-g} dx^4,
\end{equation}\\
where the function $ f(R)=R+\alpha R^2+\beta R^3 $ is used for inflation, which was first suggested by Starobinsky \cite{ref16}. The function $ f(R) $ can be treated as a quantum correction in General Relativity (GR), where $ R^2 $
produces an inflationary scenario in the early times. The Starobinsky model is most suitable model according as the latest observations \cite{pla} and may be studied to describe inflation in addition to scalar field models \cite{jdb} . Clearly, the function $ f(R) $ containing negative powers of $ R $ is capable to explain the current accelerating expansion of the Universe.\\

\qquad The Einstein Field Equation (EFE) in $ f(R) $-gravity are given as:

\begin{equation}{\label{1}}
f'(R)R_{ij}-\frac{1}{2}f(R)g_{ij}= T_{ij} -g_{ij}\square f'(R)+\nabla_i\nabla_jf'(R),
\end{equation}\\
where we have taken $ 8 \pi G $ and velocity of light c as unity. In this equation $ f'(R) $ is the differentiation of $ f(R) $ with respect to $ R $ and other mathematical symbols have their usual meanings.\\ 
 
 We consider $ FLRW $ flat space-time, which is a SHI filled with perfect fluid is given by  
 \begin{equation}{\label{2}}
 ds^2=dt^2-a^2(t)(dx^2+dy^2+dz^2).
 \end{equation}
 The energy momentum tensor of  perfect fluid  is as follows
 \begin{equation}\label{3}
 T_{ij}=(\rho_m +p_m)u_{i}u_{j}-p g_{ij},
 \end{equation}
 where $ \rho_m $ and $ p_m $ are the matter density and the pressure of  perfect fluid. We use co moving coordinates so that the four velocity $ u^i $ satisfies $ u^i u_i = 1 $.\\
 
 The field equations given by Eq.(\ref{1}) are solved for metric Eq. (\ref{2}) and energy momentum tensor given by Eq.(\ref{3}) are obtained as
 \begin{equation}{\label{4}}
 3\frac{\dot {a^2}}{a^2}=\frac{\rho_m}{f'}+\frac{1}{f'} \bigg[\frac{1}{2}(f-f'R)-3\frac{\dot{a}}{a}\dot{R}f'' \bigg]
 \end{equation}
 and
 \begin{equation}{\label{5}}
 2\frac{\ddot{a}}{a}+\frac{\dot {a^2}}{a^2}=-\frac{1}{f'} \bigg[3\frac{\dot{a}}{a}f''\dot{R}+f'''\dot R^2+f''\ddot{R}+\frac{Rf'}{2}-\frac{f}{2} \bigg],
 \end{equation}
 where prime and dot stand for derivative \textit{w.r.t.} $ R $ and proper time $ t $ respectively. At present,  we take pressure-less fluid as $ p_m \simeq 0 $ .\\
 
 The Field Eqns. (\ref{4}) and  (\ref{5}) becomes Einstein Field Eqns. for FLRW space-time when $ f(R)= R $. The additional expressions appearing in these Eqns. on the RHS are present due to non-linear function f(R). We associate them to curvature pressure $ p_k $ and curvature energy $ \rho_k $ as follows:
 
  \begin{equation}{\label{6}}
  \rho_{k}=\frac{1}{2}(f-R f')-3\frac{\dot{a}}{a}\dot{R}f''
  \end{equation}
  and
  \begin{equation}{\label{7}}
  p_k=3\frac{\dot{a}}{a}f''\dot{R}+f'''\dot R^2+f''\ddot{R}+\frac{Rf'}{2}-\frac{f}{2}.
  \end{equation}
  
  More specifically, we can express Field Eqns. (\ref{4}) and  (\ref{5}) in terms of the Hubble parameter $ H=\frac{\dot{a}}{a} $ and its derivative $\dot{H}$ as follows: 
  
  \begin{equation}{\label{8}}
  3H^2 = \frac{\rho_m +\rho_k}{f'},
  \end{equation}
  and
  \begin{equation}{\label{9}}
  2\dot{H}+3H^2=-\frac{p_k}{f'}.
  \end{equation}
   
  The purpose of this work is to form a cosmological model of the universe which fits best on the observational ground. At present, we have two types of data sets which are described as follows:
   \begin{itemize}
   	\item {\bf  Hubble Observational Data Set (OHD)}: A data set of $ 46 $ observational points of Hubble parameters  in the redshift range $ 0\leq z\leq 2.36 $, which are procured from cosmic chronometric technique (CCT) \cite{ref27}.\\
   	
   	\item {\bf Pantheon data Set}: A data set of  $ 1048 $ SNIa apparent magnitude measurements known as the Pantheon compilation \cite{ref28} includes $ 276 $ SNIa data set of the Pan-STARRS1 Medium Deep Survey in the redshift range $ 0.03 < z < 0.65 $ along with SNIa data set from SDSS, SNLS and low-$ z $ HST samples.\\
   \end{itemize}

  Now, we express the field Equations (\ref{8}) and (\ref{9}) in terms of derivatives of the $ H $ \textit{w.e.t.} redshift $ z $. For this we use the following transformation formula for redshift $ 1+z=\frac{a_0}{a} $, where $ a_0 $ is the present value of $ a $. The following field equations are obtained as per our requirement:
  
  \begin{multline}
  \label{10}
  3 f' H^2 + 0.5 \bigg(f +6 f'(2 H^2 -(1+z) H \frac{d H}{dz} \bigg) +
  18 (1+z) f'' H^2 \bigg(3 H \frac{d H}{dz}-(1+z) \bigg(\frac{d H}{dz}\bigg)^2 - (1+z) H \frac{d^2 H}{dz^2} \bigg)=\rho_m,
  \end{multline}
  
  \begin{multline}
  \label{11}
  \frac{d^3 H}{d z^3}=\frac{3Hf' - 0.5 f - 6 f'H^2}{6(1 + z)^3 f'' H^3} +
  \frac{dH}{dz} \bigg(\frac{f' + 36 f''H^2 -  6(1 + z)  f'' H^2}{6(1 + z)^2 f'' H^2}  \bigg) +  
  \left( \frac{dH}{dz} \right)^2  \bigg(\frac{1}{(1 + z)H} + \frac{54f'''H}{(1 + z) f''} \bigg)+ \\
  \left(\frac{dH}{dz} \right)^3  \bigg(\frac{-36 f'''}{f''} + \frac{1}{H^2} \bigg) + 
  \bigg(\frac{dH}{dz} \bigg)^4  \bigg(\frac{6(1 + z) f'''}{f''H} \bigg) + 
  \frac{d^2 H}{d z^2}  \bigg(\frac{-3 -(1+z)}{ (1 + z)^2} \bigg) \\ +
  \bigg(\frac{d^2 H}{d z^2} \bigg)^2  \bigg(\frac{ 6(1 + z) f''' H}{ f''} \bigg)+ 
  \frac{dH}{dz} \frac{d^2 H}{dz^2} \bigg(\frac{ -36f''' H^2 + 4f''}{f'' H} \bigg) +
  \bigg(\frac{dH}{dz}\bigg)^2 \frac{d^2 H}{d z^2}  \bigg(\frac{6(1 + z) f'''}{
  	f''} \bigg),
  \end{multline}
    In obtaining Eqns. (\ref{10}) and (\ref{11}), the following expressions in terms of $ R $, $\dot{R}$, $ H $ and $ \frac{dH}{dz}$ are used:
    
    \begin{align}\label{12}
    R &= -6(\dot{H}+ 2 H^2)\\
      &=6H\left(\frac{dH}{dz}(1+z)-2H\right),
    \end{align}
    \begin{align}\label{13}
    \dot{R} &= -6( \ddot{H} +4 H\dot{H})\\
    &= -6\left(-3H^2(1+z)\frac{dH}{dz}+(1+z)^2H\left(\frac{dH}{dz}\right)^2+H^2(1+z)^2\frac{d^2H}{dz^2}\right),
    \end{align}
    and
    \begin{multline}\label{14}
    \ddot{R}= -24 (1+z)^2 H^2 \bigg(\frac{d H}{d z} \bigg)^2\\  -18 (1+z) H^3 \frac{d H}{d z} -6 (1+z)^2 H^3 \frac{d^2 H}{d z^2}-24 (1+z)^3 H^2  \bigg( \frac{d H}{d z} \bigg) \bigg( \frac{d^2 H}{d z^2} \bigg) + 6 (1+z)^3 H \bigg( \frac{d H}{dz} \bigg)^3 + 6 (1+z)^3 H^3 \frac{d^3 H}{d z^3}.
    \end{multline}  
    
 \section{ Numerical solutions and the Dynamics of the Model} 
 \qquad As stated earlier, we want to develop a model which provides an acceleration in the universe at present. For this purpose, we consider a particular form $ f(R)= R + \alpha R^2 +\beta R^3 $. Earlier, some authors \cite{ref15, ref16} have used second order non-linear Ricci scalar to develope an accelerating universe model. So, we got motivation from the earlier work \cite{ref19, ref20} who have used $ f(R) $-gravity in developing their model.  So, $ f'(R)=1+2\alpha R+3\beta R^2 $, $ f''(R)= 2\alpha + 6\beta R $ and $ f'''(R)= 6\beta $.
 
 \subsubsection{ The deceleration and jerk parameters} 
 \qquad The deceleration parameter $ q $ and jerk parameter $ j $ in terms of $ a $ and $ H $ are defined as: 
  \begin{align*}
 q &= -\frac{\ddot{a}}{aH^2} \\
 &= -1 + (1+z) \frac{1}{H} \frac{dH}{dz},
 \end{align*}
 and 
 \begin{align*}
 j &=\frac{\dddot{a}}{aH^3}\\
 &=1+3\frac{\dot{H}}{H^2}+\frac{\ddot{H}}{H^3}\\
 &=1-2\frac{(1+z)}{H} \frac{dH}{dz}+ \frac{(1+z)^2}{H^2} \left(\frac{dH}{dz}\right)^2 + \frac{(1+z)^2}{H}\frac{d^2 H}{dz^2}.
 \end{align*}
 
The present values of the deceleration parameter $ q_0=-0.55 $ and jerk parameter $ j_0=1 $ of the obtained model is consistent with $ \Lambda $CDM model. Planck's latest survey \cite{ref27} provides us the present value of Hubble constant $ H_0\simeq 68\, Mpc/Km/sec\simeq 0.07\, Gyr^{-1} $. From these values, we can estimate the present values of first and second derivatives of Hubble parameter as $ (\frac{dH}{dz})_0 = 0.0315 $ and $ (\frac{d^2 H}{dz^2})_0 = 0.48825 $.\\
 
 The energy parameter $ \Omega_m $ is defined as $ \Omega_m=\frac{\rho_m}{\rho_c} $, where 
 $ \rho_c= \frac{8\pi G}{3 H^2} $ is the critical density. Its present value is given as $ \rho_{c0} = 1.88 \times 10^{-29}h_0^2 \, gm/cm^3 $, where $ 0.5<h_0<1 $. EoS-parameter for curvature inspired pressure and density may be defined as $\omega_k=\frac{p_k}{{\rho}_k}$. Since $ \rho_{c0} $ is negligibly small, so we shall make certain corrections as $ \rho_{m0} = \Omega_{m0} $ for further calculations. If we put the present values of $ H $ and its derivatives in Eqns. (\ref{6}),(\ref{7}) and (\ref{10}), we get the following relations amongst the parameters $\alpha,\beta, \Omega_{m0}$ and $\omega_{k0}$ as:
  \begin{equation}{\label{17}}
  \beta =261.437+1.55647~\alpha-17784.8~ \Omega_{m0}
  \end{equation}
  \begin{equation}{\label{18}}     
  \alpha = \frac{22.5937+32.2766~\omega_{k0} -952.293~ \Omega_{m0}-2195.69~ \omega_k ~\Omega_{m0}}{0.7+ \omega_{k0}}
  \end{equation}
  
\begin{figure}[H]
\begin{center}
$%
\begin{array}{c@{\hspace{0.1in}}cc}
\includegraphics[width=3.0 in, height=2.5 in]{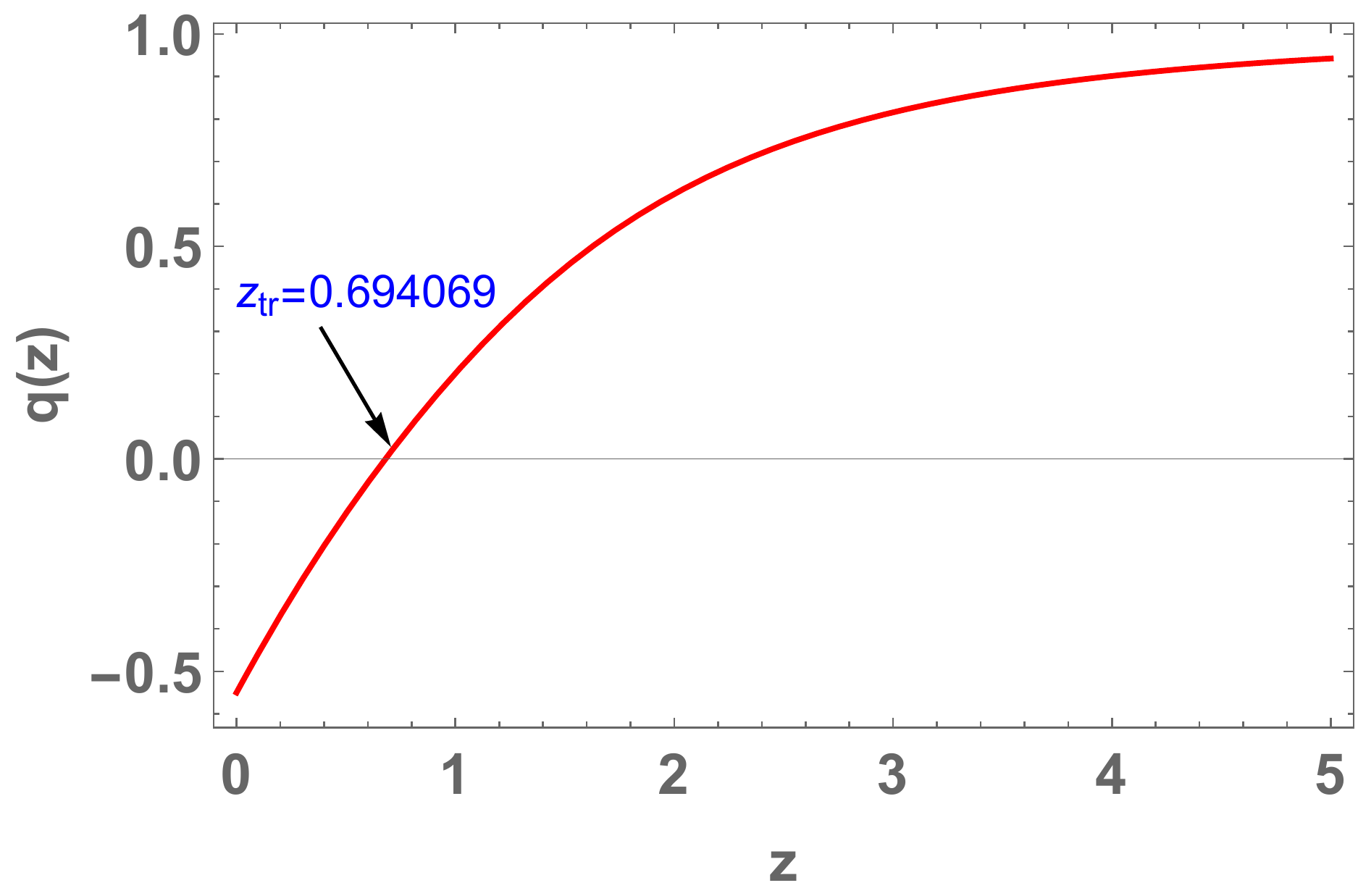} & 
\includegraphics[width=3.0 in, height=2.5 in]{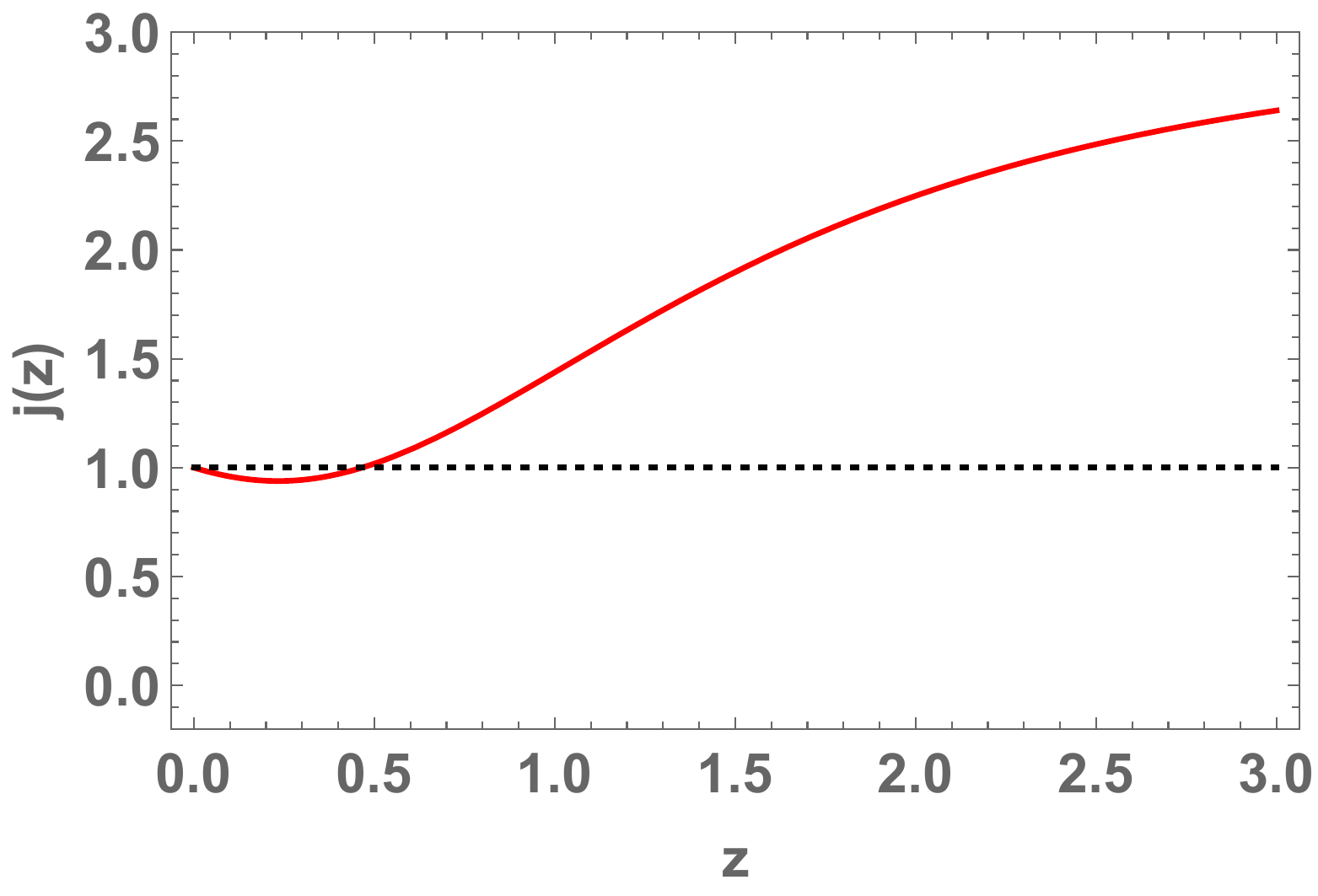}\\ 
\mbox (a) & \mbox (b)%
\end{array}%
$%
\end{center}
\caption{\scriptsize The plots of $ q-z $ and $ j-z $.}
\end{figure}     
  
In Fig. 2, we see that the deceleration parameter $ q $ is negative at present. This shows that we are living in an accelerating universe. The red shift transition is shifted from positive to negative values at $ z_{tr}\simeq 0.694069 $, which shows that the universe transforms from decelerating mode in the past to accelerating mode in late times. The jerk parameter approaches to $ \Lambda $CDM because $ j\rightarrow 1 $ as $ z\rightarrow 0 $. 
 
\subsubsection{ Statistical Analysis of the parameter} 
\qquad As stated earlier in Section(II), we have Hubble parameters data set of $ 46 $ points indicated as $ H_{obs} $.  Since Eqn. (\ref{11}) is a  non-linear third order differential equation in derivatives of $ H $ and it is not possible to find its analytical solution, therefore, we go for getting its numerical solution by putting initial values of first and second derivatives of $ H $. We call the corresponding values of Hubble parameter obtained on the basis of Eqn. (\ref{11}) as $ H_{th} $.\\

Now, we use Chi-square test to get better approximations of the parameters $ \Omega_{m0} $ and $ \omega_{k0} $. It is defined as
   \begin{equation}\label{chi}
   \chi^{2}=\sum\limits_{i=1}^{46}\frac{[H_{obs}(z_{i})-H_{th}(z_{i})]^{2}}{err {(z_{i})}^{2}},
   \end{equation}
where $ err{(z_{i})} $ stands for the standard error in OHD.\\
    
   The minimum value of Chi-square is found to be $ \chi^2_{min} =19.348 $ for $ \omega_{k0} = -0.44 $ and $ \Omega_{m0} =0.25 $. From the eqns. (\ref{17}) and (\ref{18}), we get $ \alpha=1.0027 $ and  $ \beta=-5074.44 $. The process is shown in the Fig. 2. in which blue dot is the estimated point $ (-0.44,0.25,19.348) $ for minimum Chi-square.
 
\begin{figure}[H]
\begin{center}
$%
\begin{array}{c@{\hspace{0.1in}}c}
\includegraphics[width=3.0 in, height=2.5 in]{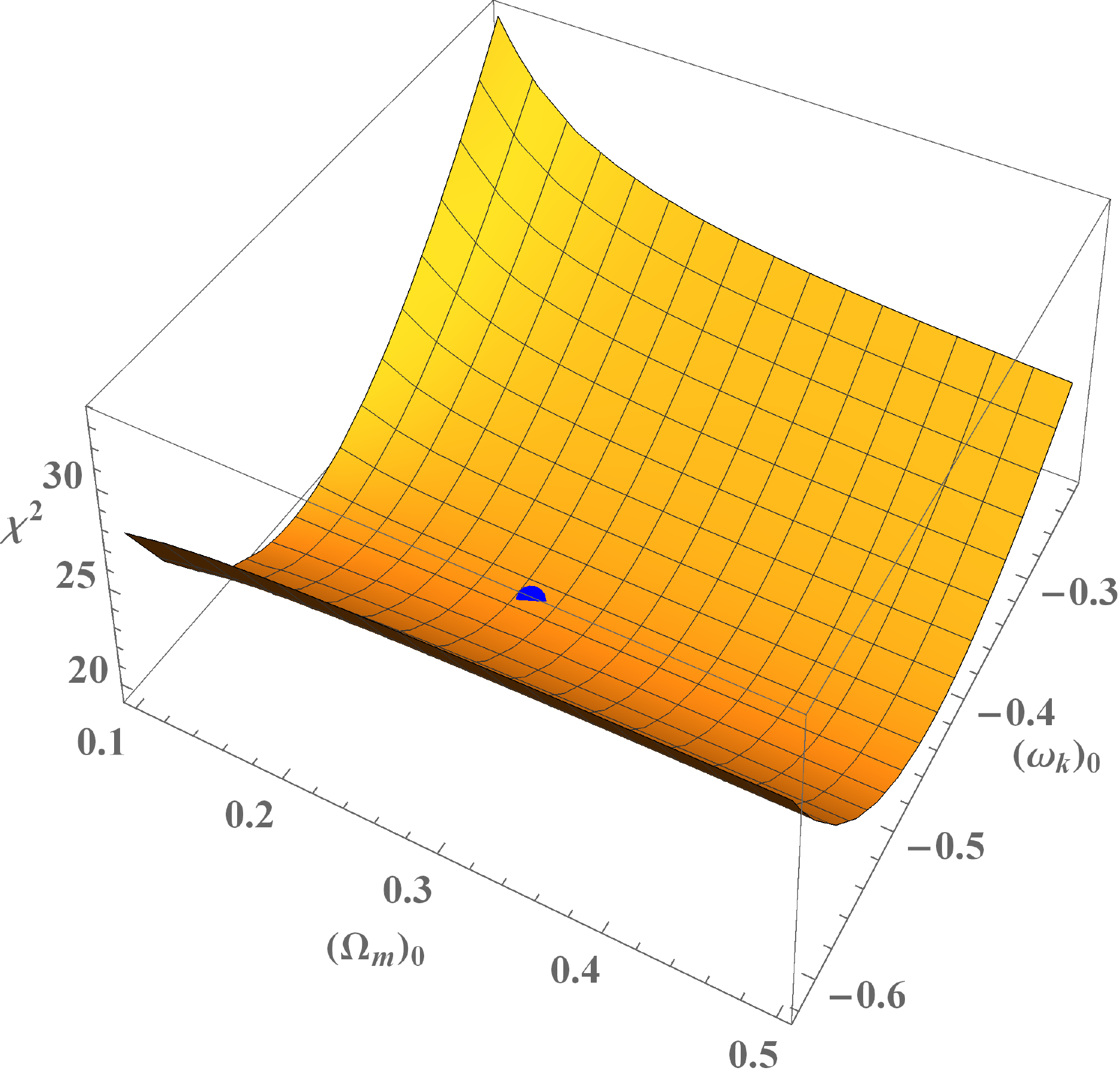} & 
\includegraphics[width=3.0 in, height=2.5 in]{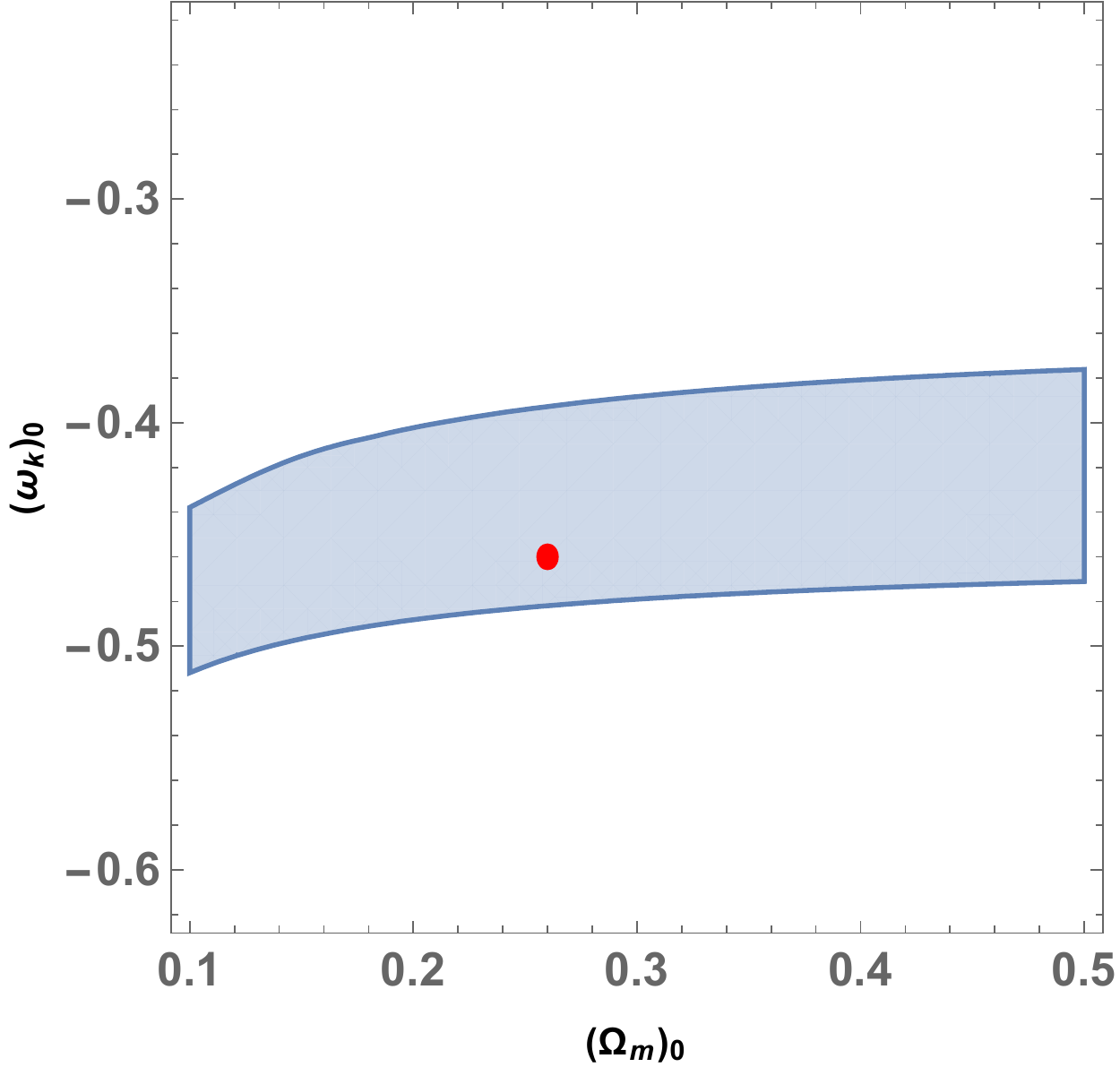}\\ 
\mbox (a) & \mbox (b)%
\end{array}%
$%
\end{center}
\caption{\scriptsize Blue and Red dots indicate the present values of $ \Omega_{m0} $ and $ \omega_{k0} $ of our model  in the Chi-square and Region plots respectively.}
\end{figure}   
 
\subsubsection{ SNIa Supernova distance modulus and apparent magnitude:} 
The luminosity distance of any  object is defined as:
\begin{equation*}
D_l(z)=(1+z)H_0 \int_0^z{\frac{1}{H(z*)} dz*},
\end{equation*}
and the distance modulus $ \mu(z) $ of a luminous object is correlated to the luminosity distance $ D_l(z) $ through the following equation as:
\begin{equation}\label{oc3}
\mu(z)= m-M = 5 Log D_l(z)+\mu_{0},
\end{equation}
where $ m $ and $ M $ are the apparent and absolute magnitude of the object and $ \mu_0= 25+5 Log \left\lbrace \frac{c}{H_0} \right\rbrace $.\\

By carrying numerical integration of tabulated values of the Hubble parameter, we have been able to find a plot $ \mu(z)-z $ for our model. We have an error bar plot for $ 620 $ supernova data sets which include $ 580 $ SNIa union compilation plus $ 40 $ binned data form \cite{ref29, ref21}. It is interesting to see that our theoretical plot coincides with the corresponding $\Lambda$CDM plot, which is displayed in Fig 2b.\\ 
 
We have presented the Hubble Error and the distance modulus plots versus red shift in Fig.2. Our plots are matching with graph prepared on the basis of $\Lambda$CDM model, which shows that these are in a good agreement with our theoretical and observed results. 
   
\begin{figure}[H]
\begin{center}
$%
\begin{array}{c@{\hspace{0.1in}}c}
\includegraphics[width=3.0 in, height=2.5 in]{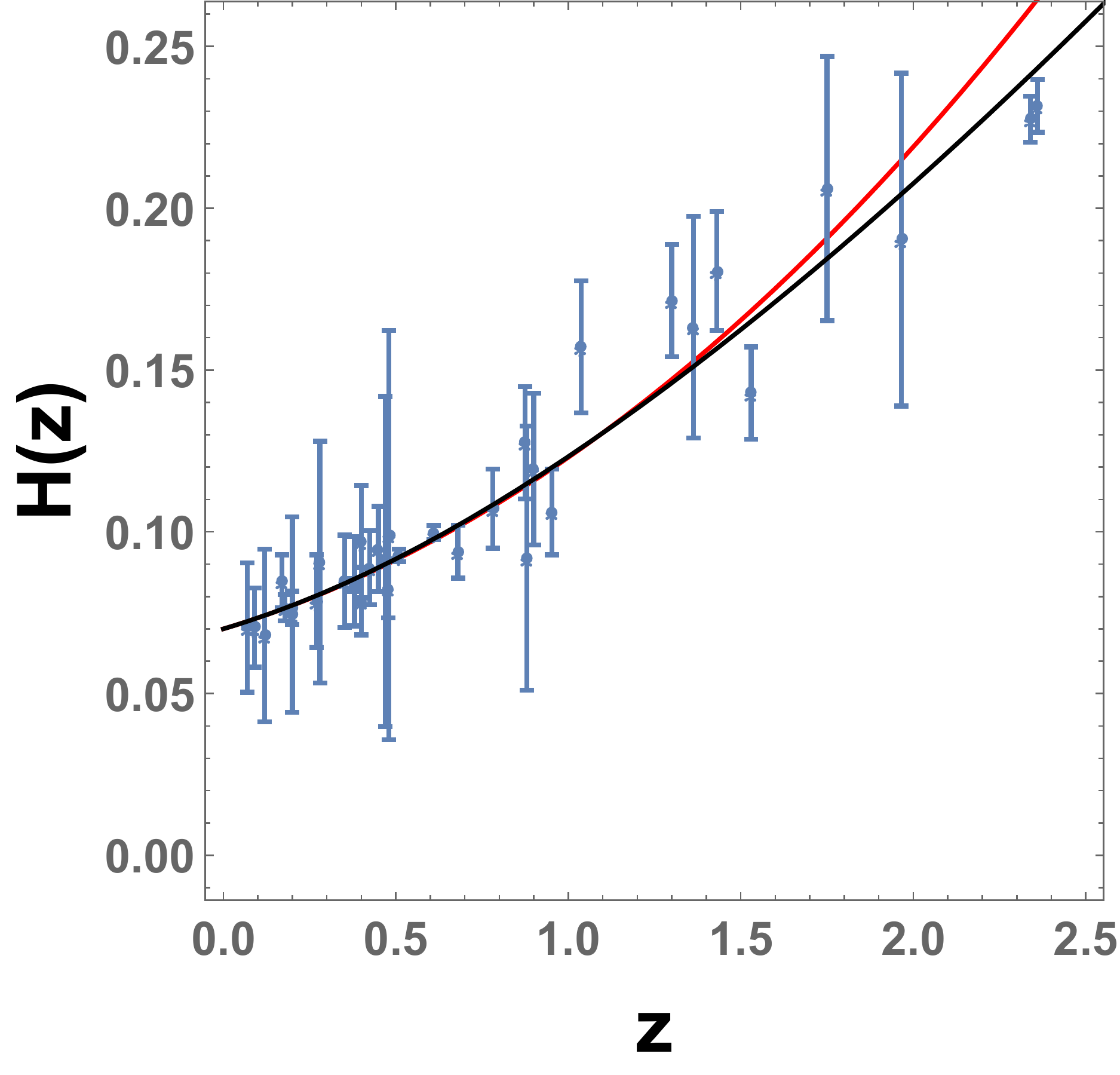} & 
\includegraphics[width=3.0 in, height=2.5 in]{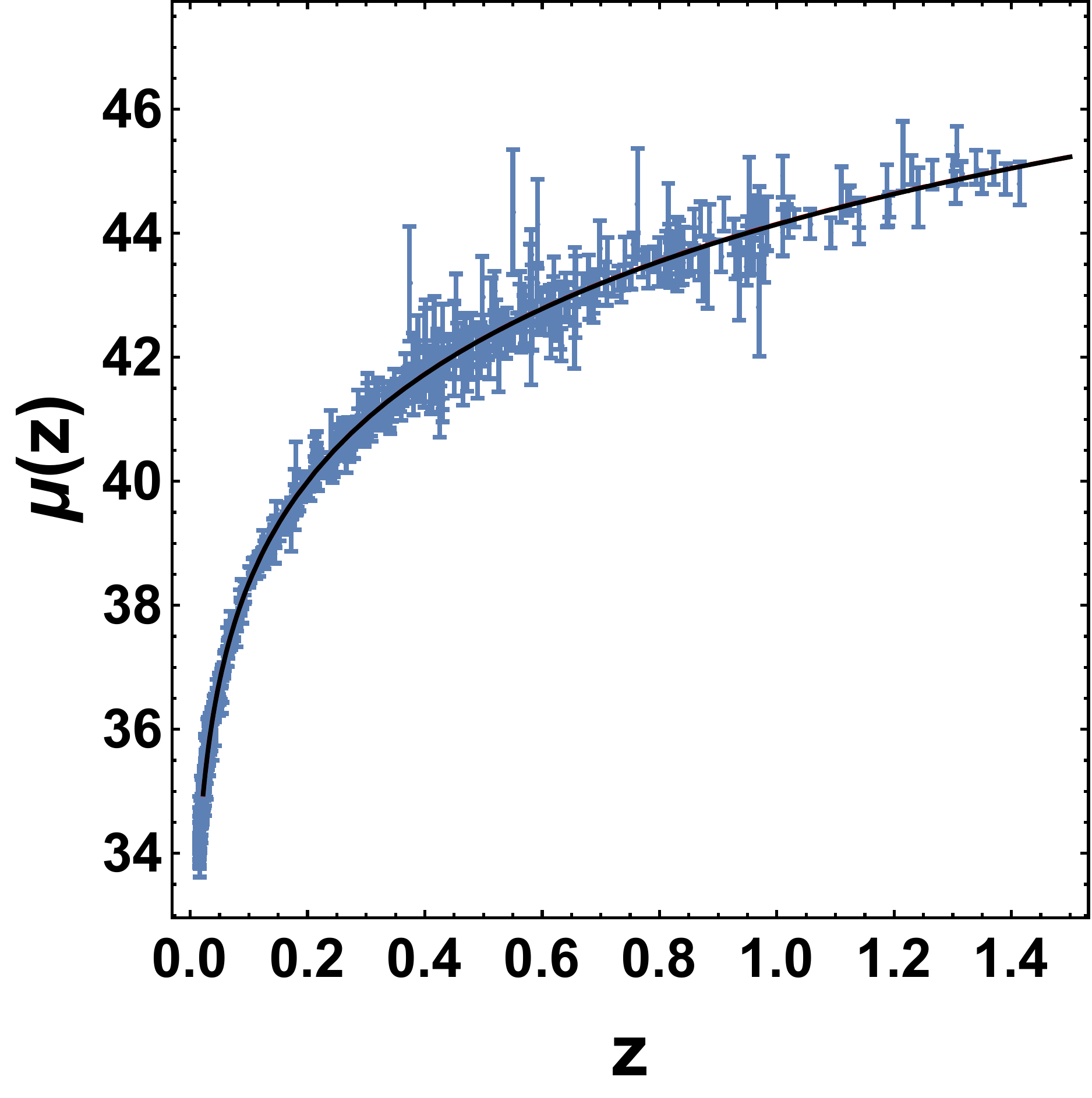}\\ 
\mbox (a) & \mbox (b)%
\end{array}%
$%
\end{center}
\caption{\scriptsize The blue color error bars of observational data $ H(z) $ and $ SNIa $ in the plots indicate the deviations of the model from $ \Lambda CDM $. Red and black lines represent our obtained model and the $ \Lambda CDM $ respectively.}
\end{figure}   

\subsubsection{ EoS-parameter and Matter energy density}
\qquad Equation of state parameter $ \omega_k $ for curvature dominated energy is defined as  $ \omega_k=\frac{p_k}{{\rho}_k} $, where $ \rho_k $ and $ p_k $ are given by Eqns. (\ref{6}) and (\ref{7}). In $ \omega_k \sim z $ plot, $ \omega_k $ is negative and is equal to $ -0.44 $ at present (see Fig. 3). The curvature dominated pressure $ p_k $ becomes negative from positive at the redshift transition  $ z \simeq 1.65 $. Therefore, we say that the accelerated dark energy model is developed due to negative pressure.\\
 
\begin{figure}[H]
\begin{center}
$%
\begin{array}{c@{\hspace{0.1in}}cc}
\includegraphics[width=3.0 in, height=2.5 in]{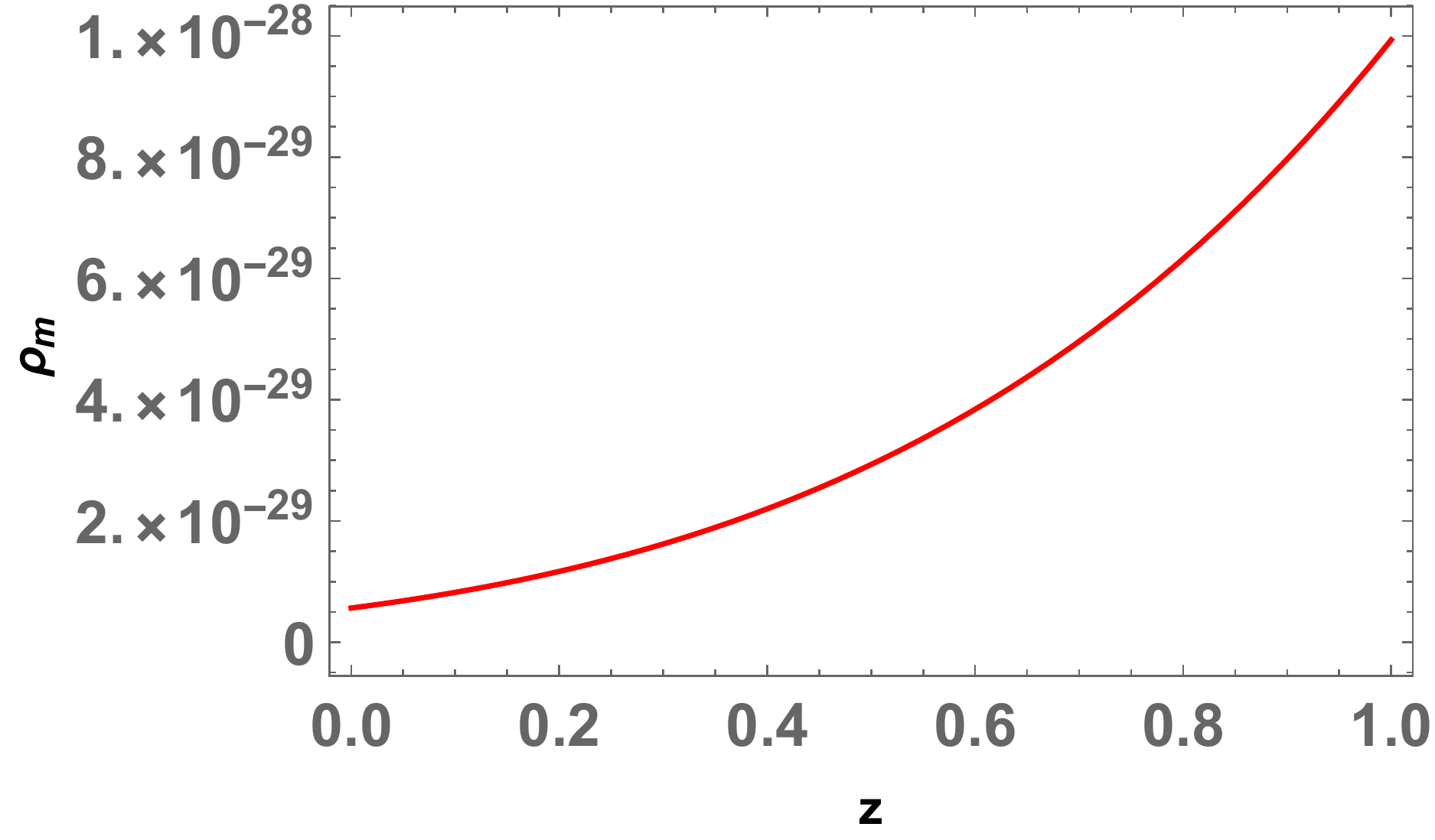} & 
\includegraphics[width=3.0 in, height=2.5 in]{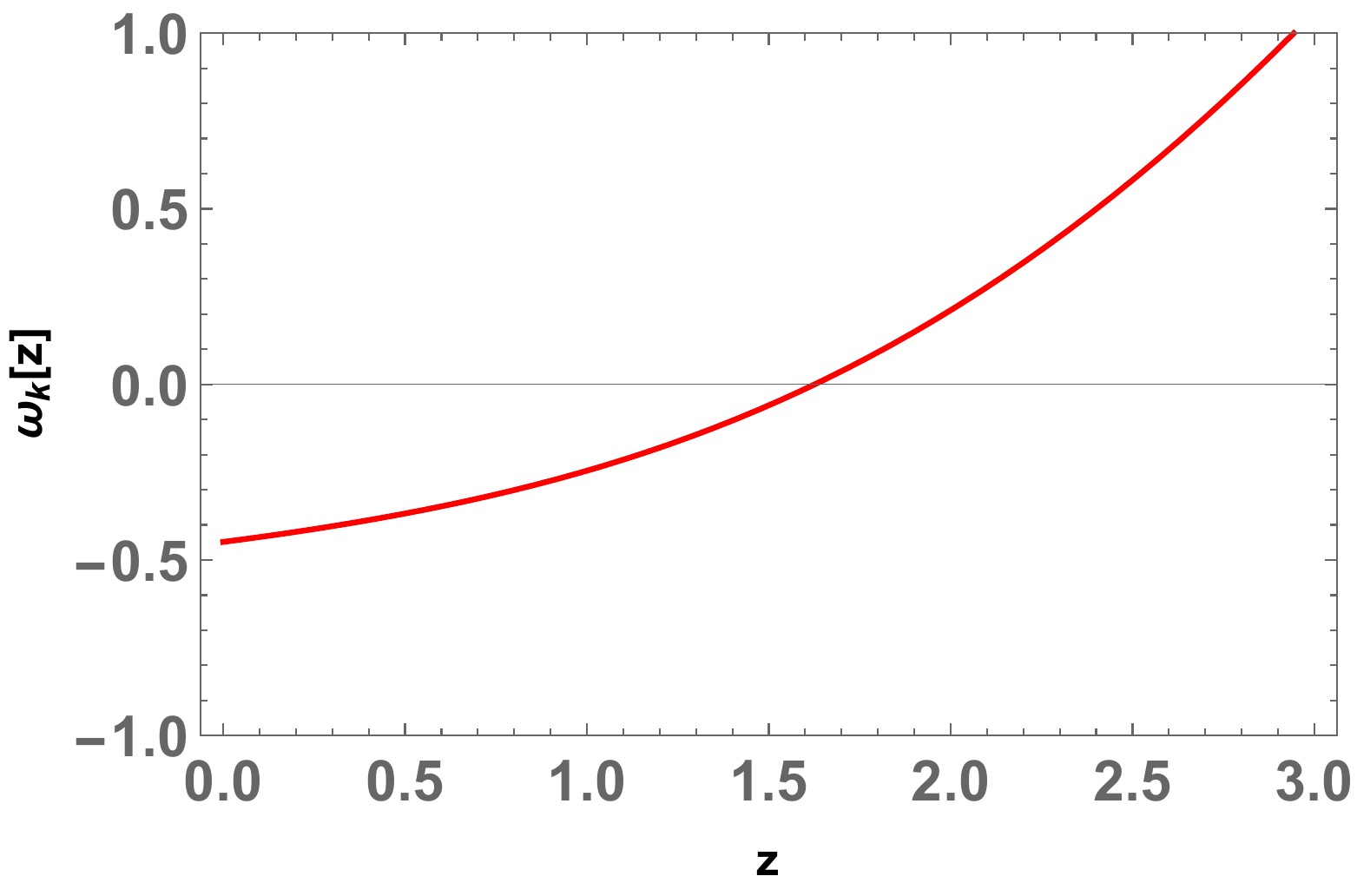}\\ 
\mbox (a) & \mbox (b)%
\end{array}%
$%
\end{center}
\caption{\scriptsize The plots of $ \rho_m-z $ and $ \omega_k-z $ .}
\end{figure}    

The numerical values of $ H $, $ \frac{dH}{dz} $ and $ \frac{d^2 H}{dz^2} $ for redshifts $ z $ are calculated from (\ref{11}) and then using (\ref{10}), we evaluate matter density $ \rho_m $. The related plot of $ \rho_m\sim z  $ can be seen in Fig. 4. As per observations, the value of matter energy density should be $ \rho_{m0}= \Omega_{m0} \times \rho_{c0} $ at present. So putting $ \Omega_{m0}\simeq 0.3 $, $ \rho_{c0}= 1.88 \times 10^{-29}\times h_0^2 \, gm/cm^3 $ and $ h_0\simeq 0.68 $ we get $ \rho_{m0}\simeq 0.27744\times10^{-29} $, $ gm/cm^3 $, which is consistent with recent observations.  

\subsubsection{State finder analysis}
\qquad In recent years, various observations have certified that our universe is accelerating at the present epoch, which caused a huge list of various DE-models have been proposed so far. Based on EoS-parameter $ \omega_{de} $ for dark energy, the most eminent models $ \Lambda $CDM ($\omega_{de}$=1), $  QCDM $ ($ \omega_{de} \geq -1/3 \leq-1 $), $ KCDM $ ($ \omega_{de} $ is negative and variable), Chaplygin gas energy model and Brain world model \textit{etc.} are proposed so far. The references related to these models can be seen in \cite{ref30, ref31}. In the year $ 2003 $, Sahni et al. \cite{ref25, ref26} proposed an analysis to examine the different category of established dark energy models. This analysis is based on deceleration parameter $ q $ and jerk parameter $ r $ as defined earlier. They have introduced a new parameter $ s $, which is defined as $ s=\frac{r-1}{3(q-1/2)},\,q \neq 1/2 $, and focused on two-dimensional $ s \sim r $ and $ q\sim r $  plots. We have presented the plots for our model (see Fig. 5). It is interesting to see that our model approaches to $ \Lambda $CDM from both ends i.e. from Chaplygin gas to  $ \Lambda $CDM and from quintessence to $ \Lambda $CDM in $ s \sim r $ plot. In $ q\sim r $ plot, our model passes nearby  $ \Lambda $CDM. So we can say that our model lies near $ \Lambda $CDM, and it fits well on observational ground.\\

\begin{figure}[H]
\begin{center}
$%
\begin{array}{c@{\hspace{0.1in}}c}
\includegraphics[width=3.0 in, height=2.5 in]{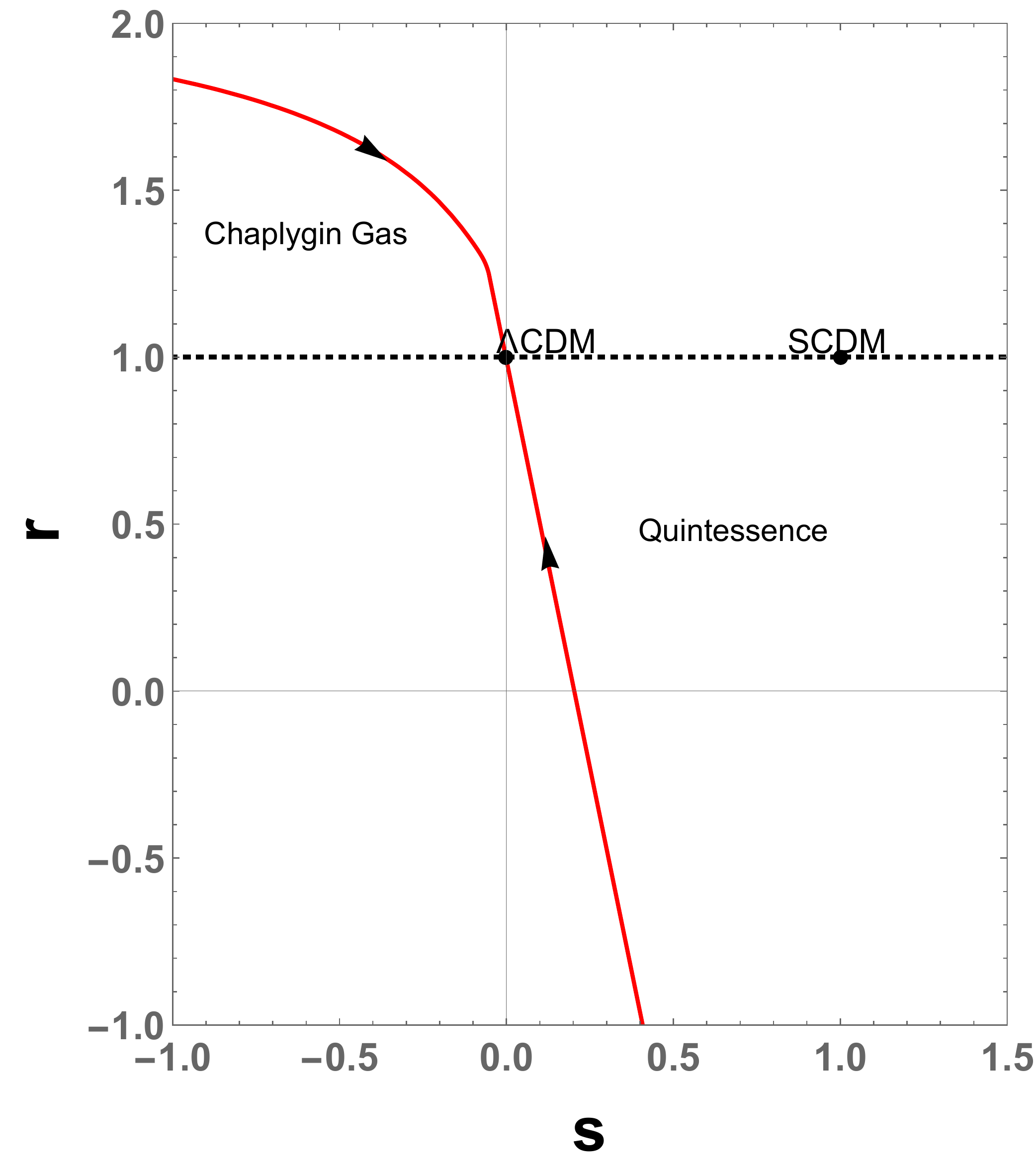} & 
\includegraphics[width=3.0 in, height=2.5 in]{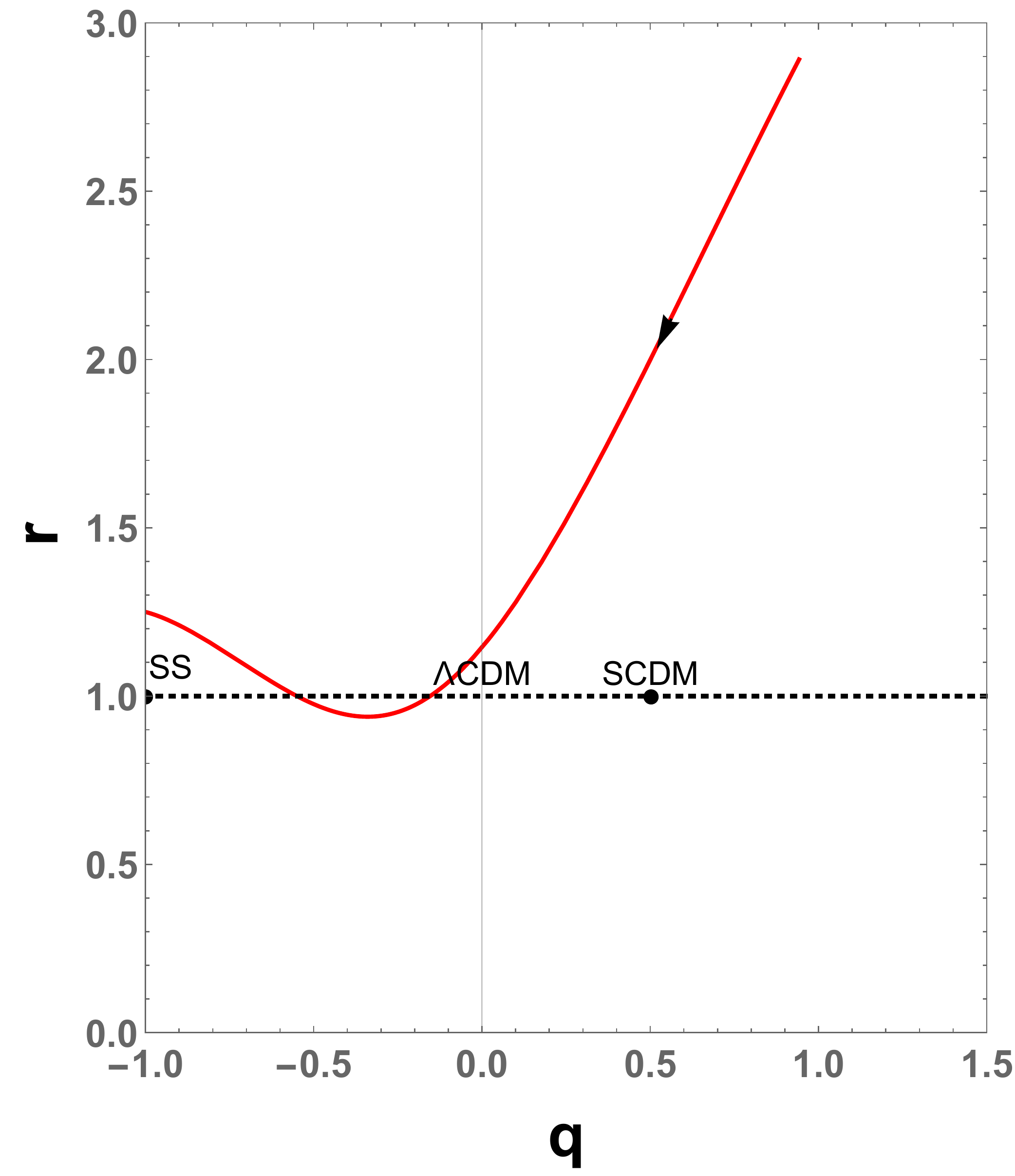}\\ 
\mbox (a) & \mbox (b)%
\end{array}%
$%
\end{center}
\caption{\scriptsize The plots of $ s-r $ and $ q-r $ for state finder analysis.}
\end{figure} 
 
\section{conclusion} 
\qquad In this work, we have probed a cosmological model with the perfect fluid filled universe in $ f(R) $-gravity  by taking $ R + a R^2 + {b}{R^3} $ as a particular form of $ f(R) $ function. The field equations (EFE) in $ f(R) $-gravity are solved for a flat FLRW spatially homogeneous and isotropic space-time. The terms which arise due to non-linear $ f(R) $ function are shifted to the RHS of the field equations that are treated as energy due to curvature inspired geometry. As a result, it produces acceleration and anti-gravitating negative pressure in the universe. The plots show that our theoretical plots fit well with the observational Hubble and Pantheon data. Moreover, the plots match at par with the $ \Lambda $CDM model. The $ q-z $ plot describes an accelerating universe at the present epoch. The transition redshift $ z $ for our model is obtained as $ z_t\simeq 0.694069 $, which is in good agreement with $ \Lambda $CDM. We have also carried out the state finder analysis for our model, which confirms that our model lies near $ \Lambda $CDM, and it fits well on the observational ground.


\begin{thebibliography}{}
\bibitem {ref1} S. Perlmutter {\it et al.},  Nature {\bf 391}, 51 (1998)
\bibitem {ref2} S. Perlmutter {\it et al.},  Astrophys. J., {\bf 517}, 5 (1999)
\bibitem {ref3}	A. G. Riess {\it et al.},  Astron. J., {\bf 116}, 1009 (1998)
\bibitem {ref4} J. P. Ostriker and P. J. Steinhardt, Nature \textbf{377}, 600 (1995), and references therein.
\bibitem {ref5} M. S. Turner, G. Steigman, and L. Krauss, Phys. Rev. Lett., \textbf{52}, 2090 (1984)
\bibitem{ref6} D. N. Spergel \textit{et al.}, [WMAP Collaboration], Astrophys. J. Suppl. Ser., \textbf{170}, (2007) 377
\bibitem{ref7} M. Tegmark \textit{et al.}, [SDSS Collaboration],  Phys. Rev. D, \textbf{69 }, 103501 (2004)
\bibitem{ref8} E. J. Copeland, M. Sami and S. Tsujikawa, Int. J. Mod. Phys. D, \textbf{15}, 1753 (2006)
\bibitem{ref9}  O. Gron and S. Hervik, Einstien's General Theory of Relativity With Modern Application in Cosmology,
(New York: Springer) (2007)
\bibitem{ref10} K. Abazajian \textit{et al.}, [SDSS Collaboration], Astron. J., \textbf{128},  502 (2004)
\bibitem{ref11} V. Sahni and A. A. Starobinsky, Int. J. Mod. Phys. D, \textbf{9}, 373 (2000)
\bibitem{ref12} S. Weinberg,  Rev. Mod. Phys. \textbf{61}, 1 ( 1989)
\bibitem{ref13} P. Steinhardt, L. Wang and I. Zlatev, Phys. Rev. D, \textbf{59}, 123504 (1999)
\bibitem{ref14} V. B. Johri, Phys. Rev. D, \textbf{63} 103504 (2001)
\bibitem{ref15} S. Nojiri and S. D. Odintsov,  Phys. Rev. D, \textbf{68} 123512 (2003)
\bibitem{ref16} A. A. Starobinsky, JETP Lett., \textbf{86} 157 (2007)
\bibitem{ref17} T. P. Sotiriou and S. Liberati, J. Phys. Conf. Ser., \textbf{ 68} 012022 (2007)
\bibitem{ref18} T. P. Sotiriou and S. Liberati, Annals Phys., \textbf{322} 935 (2007)
\bibitem{ref19} S. K. Srivastava and  Phys. Lett, B, \textbf{648} (2007) 119 (2007)
\bibitem{ref20} A. Mukherjee and N. Banerjee, Astrophys. Space Sci., \textbf{352} 893 (2014)
\bibitem{ref21} G. C. Samantha and N. Godani,  Ind. J. Phys., \textbf{94} 1303 (2020)
\bibitem{ref22} Ciprian A. Sporea, arXiv:1403.3852v2 [gr-qc] (2014)
\bibitem{ref23} M. Mali and S. Shankaranarayanan, Nuclear Phys. B, \textbf{937}, 422 (2018)
\bibitem{ref24} S. Capozziello,  Int. J. Mod. Phys. D, \textbf{11}, 483 (2002)
\bibitem{ref25} Jonathan D. Evans, Lisa M. H. Hall, and Philippe Caillol, Phys. Rev. D, \textbf{77}, 083514 (2008)
\bibitem{ref26} J. D. Barrow and S. Cotsakis,  Phys. Lett. B, \textbf{214} 515518, (1988)

\bibitem{pla} P. A. R. Ade \textit{et al.}, (Planck 2015 Collaboration), A. A., \textbf{594} A20 (2016)

\bibitem{jdb} J. D. Barrow and S. Cotsakis, Phys. Lett. B, \textbf{214} 515 (1988)

\bibitem{ref27} P. Biswas, P. Roy and R. Biswas, arXiv: 1908.00408 [gr-qc] (2019)
\bibitem{ref28} D. M. Scolnic \textit{et al.}, \textit{Astrophys. J}, \textbf{859} 101 (2018)
\bibitem{ref29} P. A. R. Ade, \textit{et al.}, Planck $\Lambda$ Collaboration, Astron. Astrophys., \textbf{594} A13 (2016)
\bibitem{ref30} N. Suzuki  \textit{et al.}, Astrophys. J., \textbf{746} (2012) 85
\bibitem{ref31} V. Sahni \textit{ et al.}, JETP Lett., \textbf{77}  201 (2003)
\bibitem{ref32} U. Alam  \textit{et al.}, Mon. Not. Roy. Astron. Soc., \textbf{344} 1057 (2003) 
\end{thebibliography}
\end{document}